\documentclass{elsart}
\usepackage{graphicx}

\def\ba{\begin{eqnarray}}
\def\ea{\end{eqnarray}}

\begin{document}

\begin{frontmatter}

\title{\bf Components of the elliptic flow in 
 Pb-Pb collisions at $\sqrt{s}= 2.76$TeV% \thanksref{grant}
} 
%\thanks[grant]{}

\author[rzu,ifj]{Piotr Bo\.zek},

\address[rzu]{Institute of Physics, Rzesz\'ow University, ul. Rejtana 16, 
35-959 Rzesz\'ow, Poland}
\address[ifj]{The H. Niewodnicza\'nski Institute of Nuclear Physics, Polish Academy of Sciences, ul. Radzikowskiego 152, 31-342 Krak\'ow, Poland}

\begin{abstract} 
We calculate the elliptic flow of charged particles in Pb-Pb 
collisions at  $\sqrt{s}$ = 2.76 TeV in relativistic
 viscous hydrodynamics.  The recent data of the ALICE Collaboration 
on the elliptic flow as function of the centrality can be very well 
described 
using the hydrodynamic expansion of a fluid with a  small shear viscosity 
$\eta/s=0.08$. The elliptic flow as function of the 
transverse momentum shows systematic deviations from a hydrodynamic 
behavior in the small momenta region $p_\perp<800$MeV. It indicates 
that a non-negligible contribution of non-thermalized particles 
from jet fragmentation is present.
\end{abstract}
\end{frontmatter}
\vspace{-7mm} PACS: 25.75.-q, 25.75.Dw, 21.65.Qr

The elliptic flow measurements  in heavy-ion collisions at
 the Large Hadron Collider (LHC) have been presented \cite{Aamodt:2010pa}.
Generally, the results are similar as observed at lower energies at 
the Relativistic Heavy Ion Collider (RHIC). The elliptic flow coefficient
 $v_2(p_\perp)$  as function of the transverse momentum is 
increasing with $p_\perp$ and saturates at higher momenta.
The elliptic flow as function of the centrality of the collision 
 reflects the initial eccentricity of the fireball at each centrality.
The elliptic flow is generated in a collective expansion of the 
dense matter created in the collision,  the comparison to model 
calculations can provide valuable information on the dynamics of the 
collisions, on the equation of state and the transport coefficients of 
the dense, hot matter \cite{Ollitrault:1992,Voloshin:2008dg}.

The integrated elliptic flow as function of the collision centrality 
can be used to extract the properties of the fluid in the expanding fireball. 
Analyzes of the elliptic flow in
 heavy-ion collisions at RHIC energies indicate that the dense matter
 is an almost perfect fluid
 \cite{Luzum:2008cw,Song:2008hj}.
The ALICE data from Pb-Pb 
collisions 
at $\sqrt{s}=2.76$TeV give smaller $v_2$ 
than expected from perfect fluid hydrodynamics with 
hadronic rescattering \cite{Hirano:2010je}. The comparison of viscous
 hydrodynamic results with data 
at RHIC and LHC show no noticeable change in the shear
 viscosity coefficient with the energy 
\cite{Luzum:2010ag}. Analysis of the LHC data 
at different centralities
in terms of the Knudsen number points towards 
a small value of the shear viscosity \cite{Lacey:2010ej}. The observed elliptic
 flow can be partly  reproduced in a  hadron rescattering model with a short 
formation time
 \cite{Humanic:2010ub}.

\begin{figure}[ht]
\begin{center}
\includegraphics[width=.65\textwidth]{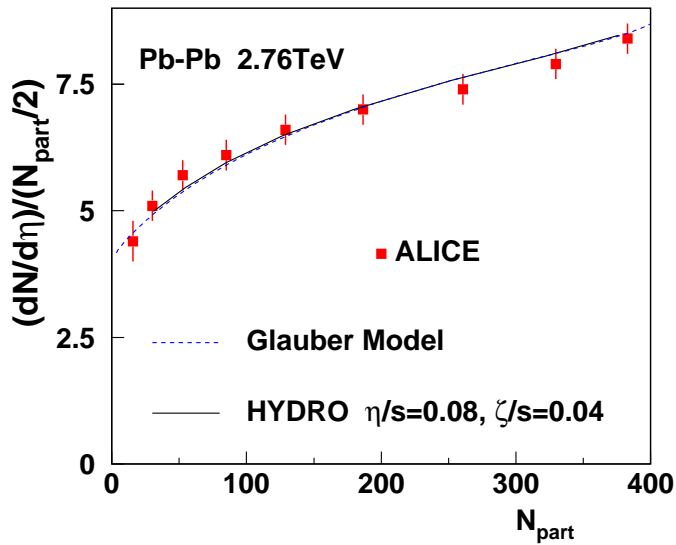}
\caption{Charged particle multiplicity density in pseudorapidity per participant pair 
for Pb-Pb collisions at $\sqrt{s}=2.76$ TeV from the Glauber
 model (dashed line), and from viscous hydrodynamic calculations 
(solid line) together with the ALICE  data \cite{Aamodt:2010cz}.}
\label{fig:dndeta}
\end{center}
\end{figure}

 In the following we present an analysis of the ALICE results 
for the charged particle elliptic flow in the second order viscous 
hydrodynamic model with shear and bulk viscosities \cite{Bozek:2009dw}.
Particle production in the central region at the LHC is described in the
$2$+$1$-dimensional, boost-invariant hydrodynamics.
The shear viscosity to entropy ratio is $\eta/s=1/4\pi$ and the bulk 
viscosity is non-zero in the hadronic phase with $\zeta/s=0.04$.
The initial entropy density of the fireball in the transverse plane 
is taken from the Glauber model, at  the impact parameter $b$
\begin{equation}
s(x,y) = s_0 \frac{(1-\alpha)\rho_{part}(x,y,b)+2\alpha
 \rho_{bin}(x,y,b)}{(1-\alpha)\rho_{part}(0,0,0)+2\alpha
 \rho_{bin}(0,0,0)} \ , 
\label{eq:den}
\end{equation}
with $\rho_{part}$ and $\rho_{bin}$ the participant nucleons 
and binary collisions densities. The details of the hydrodynamic model 
calculation and the parameters for Pb-Pb collisions at the LHC can be 
found in \cite{Bozek:2010er,Bozek:2009dw}. 
 The statistical emission at the freeze-out and resonance 
decays are done using the THERMINATOR event generator \cite{Kisiel:2005hn}.
The parameter $\alpha=0.15$ reproduces the
centrality dependence of the density of charged particles in pseudorapidity 
$\frac{dN}{d\eta}$ measured by the ALICE Collaboration 
\begin{equation}
\frac{dN}{d\eta}=\frac{dN_{pp}}{d\eta} 
\left(\frac{1-\alpha}{2}N_{part}+\alpha N_{bin}\right) 
 \end{equation}
(dashed line in Fig \ref{fig:dndeta}), where $dN_{pp}/d\eta$ is the 
multiplicity density 
in inelastic proton-proton collisions, $N_{part}$ and $N_{bin}$
are the average numbers of participants and of binary collisions for a
 given centrality.
 The entropy 
density at the center of the fireball $s_0$ in Eq. \ref{eq:den} is adjusted 
to reproduce the multiplicity in the most central Pb-Pb 
collisions, taking into 
account the entropy production in the viscous hydrodynamic evolution.
The results of the hydrodynamic calculation (solid line in 
Fig. \ref{fig:dndeta}) follow closely the Glauber model input and 
reproduce the data on the centrality dependence of the multiplicity 
\cite{Aamodt:2010cz}.

\begin{figure}[t]
\begin{center}
\includegraphics[angle=0,width=0.85\textwidth]{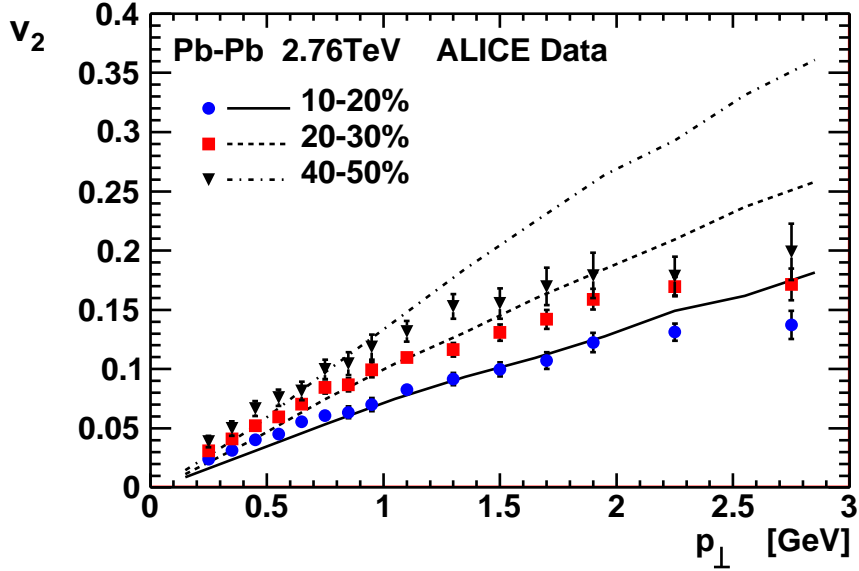}
\end{center}
\caption{\small Elliptic flow coefficient of charged particles as function 
of transverse momentum measured by the ALICE Collaboration \cite{Aamodt:2010pa}
 (symbols) together with the results of viscous hydrodynamics  (lines) for three
 centrality classes.}
\label{fig:v2}
\end{figure}

For each centrality class the elliptic flow of charged particles as 
function of the transverse momentum is calculated in the reaction plane
 (Fig. \ref{fig:v2}). For central 
 collisions ($10$-$20$\%) the results of the hydrodynamic calculation
 follow the data up to $p_\perp\simeq 2$GeV. For more peripheral
 collisions ($40$-$50$\%) the data show a saturation of
 the increase of $v_2$ with $p_\perp$ around $1.2$GeV, 
while the calculated $v_2(p_\perp)$ keeps growing with the transverse momentum.
The change in the behavior of the elliptic flow coefficient with 
$p_\perp$ is expected. The production of particles with  
small momenta $p_\perp<1$-$2$GeV is
 dominated by the statistical emission  from a thermalized fluid. 
The asymmetry of the flow of the fluid comes from the hydrodynamic 
evolution of a source with an initial azimuthal eccentricity.
 On the other hand, particles at high momenta originate from 
jet fragmentation
\cite{Wiedemann:2009sh,d'Enterria:2010zz}. Differences in the
 path length in different directions in the fireball, lead via 
jet quenching to an azimuthal asymmetry of the final particles.
While it is expected that the hydrodynamic results deviate from the data 
at high transverse momenta, we observe also a systematic deviation of the
data points from the calculation in the range $200$MeV$<p_\perp<800$MeV.
We discuss about this effect latter.

\begin{figure}[t]
\begin{center}
\includegraphics[angle=0,width=0.65\textwidth]{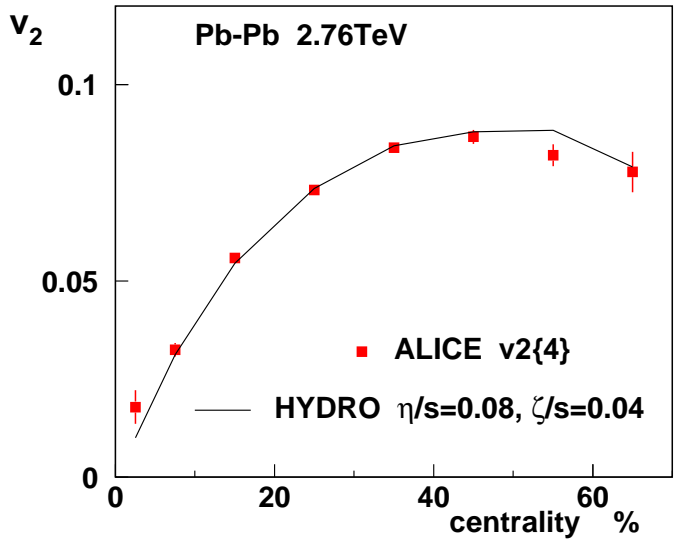}
\end{center}
\caption{\small Elliptic flow coefficient in Pb-Pb collisions 
as function of centrality, ALICE Collaboration data \cite{Aamodt:2010pa} 
compared to  viscous hydrodynamic results.}
\label{fig:v2ch}
\end{figure}

In Fig. \ref{fig:v2ch} is shown the integrated elliptic 
flow $v_2$ for different centralities of the collision  
($|\eta|<0.8$, $200$MeV$<p_\perp<5$TeV). The 
hydrodynamic model using Glauber model initial sources, with shear
 viscosity $\eta/s=0.08$ and bulk viscosity $\zeta/s=0.04$ reproduces 
the experimental observations very well. It means that, for the chosen 
initial eccentricity, the expansion of the fluid with a minimal shear viscosity 
describes the data. 
The analogous calculation for Au-Au collisions at $200$GeV describes well 
the RHIC data 
for the integrated $v_2$ using the same value $\eta/s=0.08$.
The precise value of the shear viscosity coefficient 
 compatible with the 
 data depends on the model of the initial eccentricity, 
with color glass condensate initial conditions leading to larger
 values of $\eta/s$ 
\cite{Luzum:2008cw,Song:2008hj}. 
We note that the Glauber binary collisions profiles of the initial energy 
density used
in Ref. \cite{Luzum:2010ag} result in  up to $20$\% larger initial eccentricities
in central collisions than Eq. \ref{eq:den}, which would imply a larger 
value of $\eta/s$ to reproduce the data.
 
Despite the apparent success of the hydrodynamic description of the average 
elliptic flow in Pb-Pb collisions at the LHC as presented in Fig. \ref{fig:v2ch},  the result must be taken with caution.
The differential elliptic flow coefficient $v_2(p_\perp)$ is not
 described by the hydrodynamic model, neither at high momenta, 
which is natural, nor at small momenta $p_\perp<800$MeV. If the elliptic 
flow originates from the collective flow of the fluid, the behavior at 
small momenta is linear in the transverse momentum 
(for particles with a small mass)
\begin{equation}
v_2(p_\perp)\propto p_\perp \ .
\end{equation}
Such a dependence is observed in hydrodynamic calculations 
(also including viscosity corrections) and in experiments 
on heavy-ion collisions at RHIC.

\begin{figure}[t]
\begin{center}
\includegraphics[angle=0,width=0.85\textwidth]{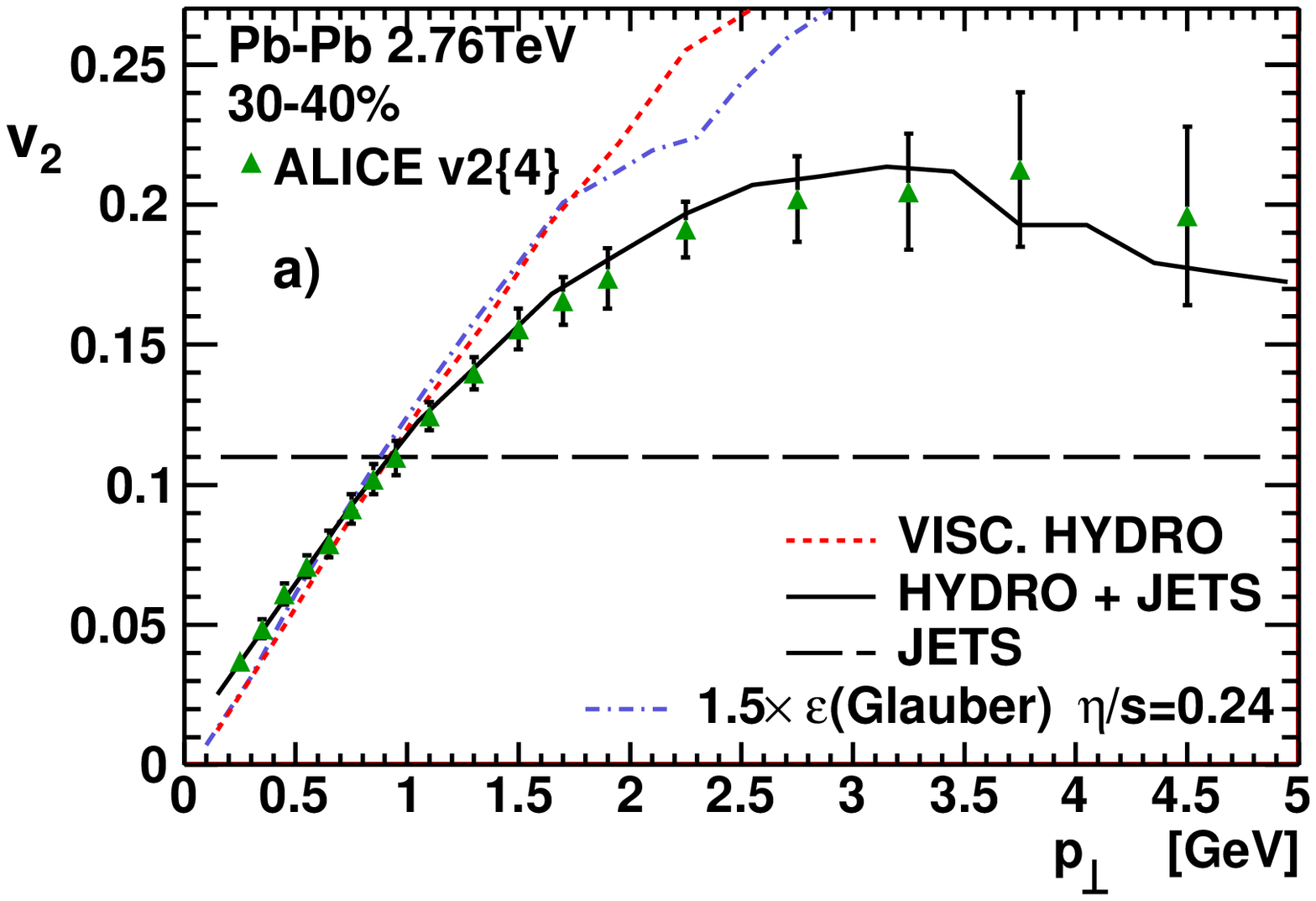}
\includegraphics[angle=0,width=0.85\textwidth]{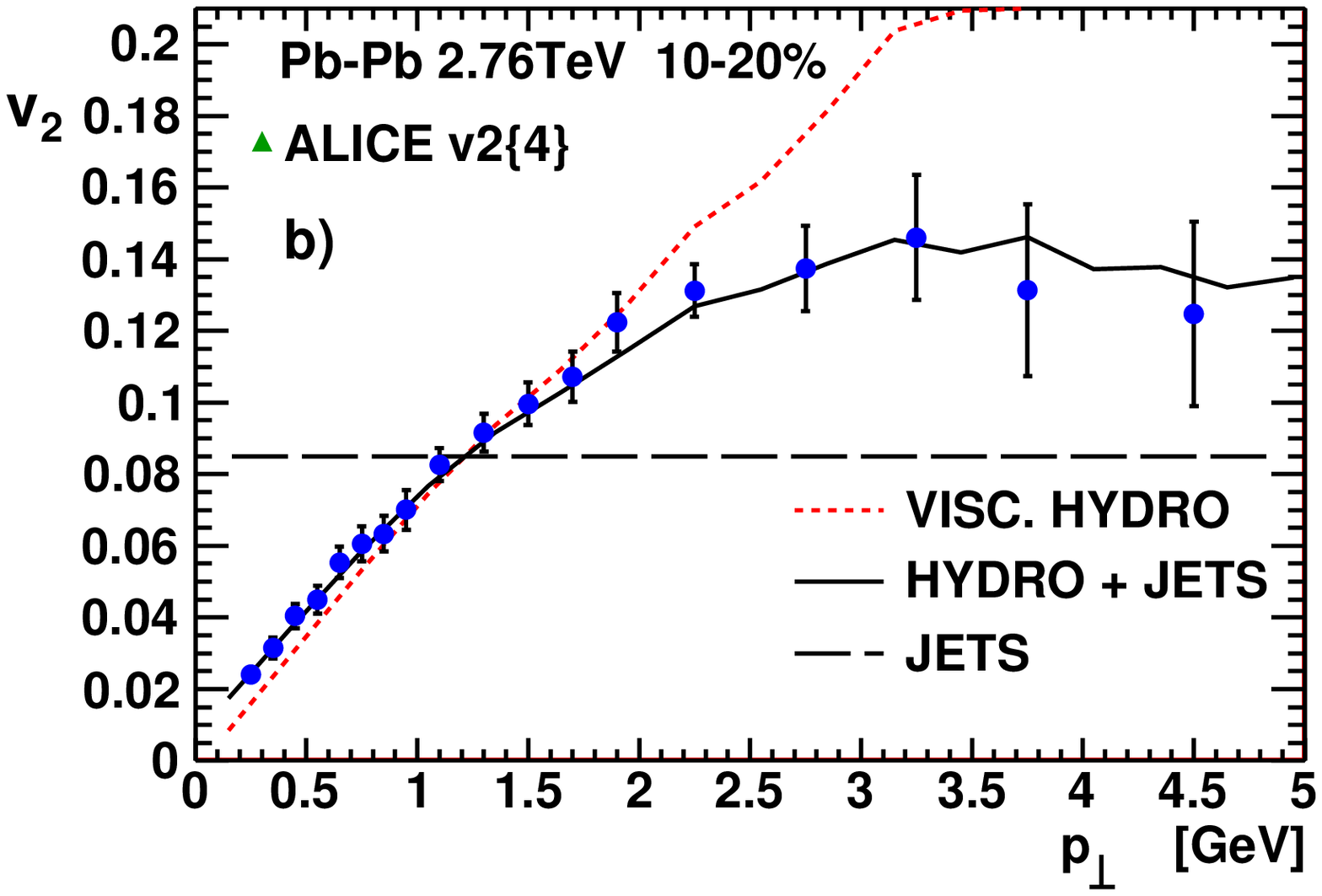}
\end{center}
\caption{\small  Elliptic flow coefficient of charged particles as function 
of transverse momentum for centrality $30$-$40$\% (upper panel) and
$10$-$20$\% (lower panel)
measured by the ALICE Collaboration \cite{Aamodt:2010pa}. The dotted line represents   viscous hydrodynamic results, the dashed line
 the approximate elliptic flow from jet fragmentation and the solid line the weighted mixture  of the jet and hydrodynamic contributions (Eq. \ref{eq:mix}).
The dashed-dotted line in the upper panel represents the elliptic flow obtained increasing the Glauber model initial eccentricity by $50\%$ and using $\eta/s=0.24$.}
\label{fig:hjet}
\end{figure}

The elliptic flow measured at the LHC is not increasing linearly with the 
momentum.
The observed differential elliptic flow $v_2(p_\perp)$ is larger 
 than in hydrodynamic calculations. The extrapolation of the measured elliptic
 flow to zero momentum is not  going through the point $v_2(0)=0$, as it 
should for 
the elliptic flow of a 
collective origin (Fig. \ref{fig:hjet}). A linear fit to the data points 
with $p_\perp<1$GeV gives an intercept of $0.06$-$0.01\pm 10^-5$,
 a small but statistically significant and systematic deviation from zero.  
Moreover, the 
extrapolation of the results of hydrodynamic calculations would lead 
to a negative intercept. It is
 due to the fact that for massive particles we have 
$v_2(p_\perp)\propto p_\perp^2 $. %The deviation between the data and 
%the calculation is of the order of $0.02$. 
While it is expected that hydrodynamic results deviate from the data 
at high momenta, it is very difficult to explain the excess of the 
observed elliptic flow  at small 
momenta $p_\perp<800$MeV.
The data points lie above the linear  function from hydrodynamic calculations
for all centralities in the range $200$MeV$<p_\perp<800$MeV. The additional 
shift of the measurements above the calculated values 
 is about
 $0.01$-$0.02$.
One possibility is that the linear behavior happens at very small momenta
 $p<200$MeV, not observed experimentally. The data points observed by 
the ALICE Collaboration ($p_\perp>200$MeV) would correspond to the 
region of strong viscous correction,  with deviations from the linear behavior.
In the upper panel of Fig. \ref{fig:hjet} is shown an example 
of a calculation
starting with initial eccentricity increased by a factor $1.5$ and
with $\eta/s=0.24$.
Such  a calculation gives the same integrated elliptic flow as observed
 experimentally. The calculated 
 elliptical flow coefficient $v_2(p_\perp)$ is closer to the data 
points, but still a systematic deviation is visible for small momenta.

The elliptic flow in hydrodynamic models originates predominantly 
 from the azimuthal 
asymmetry of the collective flow velocity. On general grounds, one
 expects a linear behavior of $v_2(p_\perp)$ for light particles at 
small momenta. There is a different mechanism generating the 
azimuthal asymmetry,
 related to the asymmetry of surface of the emission. Due to strong
jet quenching in the dense matter, the escaping partons originate from the 
surface of the fireball. In peripheral collisions the geometrical asymmetry 
of the  source leads to preferential emission of escaping partons 
in-plane. The elliptic flow coefficient of the emitted partons
would be approximately independent of the parton momentum \cite{Shuryak:2001me}.
The final observed hadrons are distributed in a cone 
around the direction of the 
fragmenting jet parton. 

At the LHC energies, in each Pb-Pb collision many jets are formed. 
The jets interact with the dense matter in the fireball, leading to jet 
quenching and jet asymmetry \cite{Aad:2010bu,Aamodt:2010jd}. The interaction 
of the jet parton 
with the matter in the fireball, depends on the path it travels, 
which leads to an azimuthal asymmetry of the particle emission at large 
transverse momenta. The elliptic flow coefficient  at large momenta at the RHIC 
energies is approximately independent of the momentum and is of the order of
$0.05$-$0.12$, depending on the centrality
\cite{afa:2009iv,Adare:2010sp}. The value of the elliptic flow
coefficients from jets at momenta $p_\perp<1$GeV is not known experimentally,
 as this
 region is dominated by 
the 
statistical emission from the
 thermalized fluid. Once
 the direction of the jet parton is fixed, the fragmenting hadrons 
 inherit, 
to a large extent, the original direction of the jet.
A particle of momentum $\simeq 200$-$800$MeV is not the leading particle in the
fragmentation of the jet, it originates as one of the soft particles in the
 fragmentation of a parton with the momentum of a few GeVs. 
The leading particle is emitted in a narrow cone
 around the jet parton. The correlation between a high momentum trigger 
particle (leading particle) and
 associated particles show that  also the soft particles are emitted
 in a cone around the jet direction. Although, one expects the jet
 cone opening is larger for particles with small momenta, we 
 assume that the elliptic 
flow coefficient from jets is approximately constant down to $200$MeV for
this first estimate.

 Following the parameterization 
of Ref.
\cite{Liao:2009ni},
we write the elliptic flow coefficient at a given momentum as the  sum of two
 components, a jet component $v^{jet}_2$ and a hydrodynamic component 
$v^{hydro}_2$
\begin{equation}
v_2(p_\perp)=(1-g(p_\perp))v^{hydro}_2(p_\perp)+g(p_\perp)v^{jet}_2(p_\perp) \ ,
\label{eq:mix}
\end{equation}
$g(p_\perp)$ denotes the proportion of particles originating
 from jet fragmentation at a given momentum. It  implicitly assumes that 
the jet particles, accounted for in the 
weight $g(p_\perp)$, are particles that do not thermalize and conserve 
their original (non-hydrodynamic) $v_2$.
There are only few constraints on the form of $v_2^{jet}$ and $g(p_\perp)$;
$g\simeq 1$ at large momenta and, from the deviation of the ALICE data 
away from the linear function $v_2(p_\perp)\propto p_\perp$, we have
$g(p_\perp)v_2^{jet}(p_\perp) \simeq 0.01$-$0.02$ for $p_\perp<800$MeV. 
For illustration, we 
fix the parameters \cite{Liao:2009ni} 
as  $g(p_\perp)=\left(1+\tanh((p-p_w)/\Delta
 p)\right)/2$, $p_w=2.8$GeV, $\Delta p=2.8$GeV, $v_{jet}(p_\perp)=0.11$ 
for centrality $30$-$40$\% and $p_w=3.2$GeV, $\Delta p=3.0$GeV, 
$v_{jet}(p_\perp)=0.085$ 
for centrality $10$-$20$\%. The final elliptic flow 
coefficient from the two components is denoted by the solid lines in Fig. 
\ref{fig:hjet}. The deviation of the data from the hydrodynamic calculation
in the range $200$MeV$<p_\perp<800$MeV
 can be explained as due to a $10$-$20$\% contribution of jet particles.
For soft particles emerging from a jet parton, one expects some reduction 
of $v_2^{jet}(p_\perp)$ for small momenta, 
due to a larger jet cone opening or to a possible 
coalescence with thermal partons. This would imply a larger value of 
$g(p_\perp)$ to get a similar $v_2^{jet}(p_\perp)g(p_\perp)$.

We present results of relativistic viscous hydrodynamic calculations for 
Pb-Pb collisions at $\sqrt{s}=2.76$TeV. The initial density for the 
evolution is fixed using the experimental results for the charged particle
 multiplicity as function of the centrality. The same calculation
 was shown  to describe
 correctly 
 the 
observed interferometry radii in central collisions at the LHC 
\cite{Bozek:2010er}. The 
hydrodynamic calculation using Glauber model initial profiles and a small shear
 viscosity coefficient reproduces the integrated elliptic flow coefficient 
observed by the ALICE Collaboration.
 Surprisingly, 
the differential elliptic flow as function 
of the transverse momentum $v_2(p_\perp)$ shows systematic deviations 
from the hydrodynamics estimate.  At small momenta the data points 
lie above the 
linear hydrodynamic behavior $v_2(p_\perp)\propto p_\perp$. It is a
 very unexpected result. In the small transverse momentum region, 
besides the dominant statistical emission from a collectively expanding 
fluid, a non-negligible contribution from jets  appears. This observation, 
if confirmed by further experimental studies, indicates that at the 
 LHC energies
non-thermalized particles from jet fragmentation constitute a substantial 
part of the soft spectrum. 
This additional source of particles with soft momenta 
should be
taken into account in quantitative studies of hydrodynamic models 
of heavy-ion collisions, in calculating the spectra and  the elliptic flow
 of particles. On the other hand, it could serve as a new frontier
 of studies of the jet parton 
formation, attenuation and fragmentation in a
 heavy-ion environment.
In particular, the fact that particles with soft momenta do not thermalize
shows that the jet fragmentation occurs {\it outside} of the thermal fireball.

{\bf Acknowledgments: }  The author thanks Raimond Snellings for providing the 
experimental data.  Research 
supported in part by the MNiSW grant No. N N202 263438.

%\bibliography{../hydr}

%\end{document}

\end{document}